# Ferromagnetism, insulator-metal transition and magnetotransport in $Pr_{0.58}Ca_{0.42}MnO_3$ films: role of microstructural perturbations

Vasudha Agarwal[a], P. K. Siwach[a], G. D. Varma[b], V. P. S. Awana[a], A. K. Srivastava[a] and H. K. Singh[a]#

[a]CSIR-National Physical Laboratory, Dr. K. S. Krishnan Marg, New Delhi-110012, India
[b]Department of Physics, Indian Institute of Technology, Roorkee, Uttarakhand-247667, India

Abstract

Magnetic and magnetotransport properties of oriented polycrystalline $Pr_{0.58}Ca_{0.42}MnO_3$ thin films prepared in flowing oxygen and air ambient has been investigated. The magnetic ground state of both the films is a frozen cluster glass. In the air annealed film charge order (CO) is quenched and ferromagnetic (FM) transition, which appears at $T_C \approx 148$ K is followed by antiferromagnetic (AFM) transition at $T_N \approx 104$ K. This film shows self-field hysteretic insulator-metal transition (IMT) at $T_{IM}^C \approx 89$ K and $T_{IM}^W \approx 148$ K in the cooling and warming cycle, respectively. Application of magnetic field (H) gradually enhances $T_{IM}^C$ and $T_{IM}^W$, reduces the thermoresistive hysteresis and $\Delta T_{IM}$ ($=T_{IM}^W - T_{IM}^C$) diminishes. In contrast, the film annealed in flowing oxygen shows a CO transition at $T_{CO} \approx 236$ K, which is followed by FM and AFM transitions at $T_C \approx 148$ K and $T_N \approx 140$ K, respectively. This film shows appreciably smaller magnetic moment and does not show IMT upto H=20 kOe. As H is increased to H=30 kOe, IMT having strong thermoresistive hysteresis and sharp resistivity jumps appears at $T_{IM}^C \approx 66$K and $T_{IM}^W \approx 144$ K in the cooling and warming cycle, respectively. As H increases to higher values the thermoresistive hysteresis is reduced, resistivity jumps are observed to disappear and $\Delta T_{IM}$ decreases. In the lower temperature regime (T=5 K and 40 K) the resistivity first decreases slowly with H and then shows sharp drop. The virgin cycle ρ is not recoverable in subsequent cycles. The ρ decrement far more pronounced in the oxygen annealed film and occurs at much higher H suggesting that the frozen cluster glass state is more robust in this film. At temperatures around $T_C$, the ρ-H hysteresis and the magnetic field induced drop in resistivity is more prominent in the oxygen annealed film. In the temperature regime dominated by CO ($T_C<T<T_{CO}$) higher H is required to induce metallic transition in the oxygen annealed film. The magnetic and magnetotransport data clearly show that the film annealed in oxygen has higher fraction of the AFM/COI phase, while the air annealed film has higher fraction of FMM phase. The microstructural analysis of the two set of films employing HRTEM clearly shows that CO quenching, FM transition and self-field IMT in air annealed film is caused by higher density of microstructural disorder and lattice defects. The difference in growth ambience of the two films could give rise to such microstructural perturbations.

# Corresponding Author; Email: hks65@nplindia.org

**Introduction**

For having a sound understanding of the physical properties of doped rare earth manganites having general formula $RE_{1-x}AE_xMnO_3$ (RE=Pr, Nd, La, etc. and AE=Ca, Sr, Ba, etc.), phase separation has emerged as the most striking intrinsic attribute [1–5]. The degree of phase separation depends on the size and relative fraction of the coexisting electronic phases, which in turn depends on the degree of interaction between fundamental variables such as the spin, lattice and charge. These fundamental degrees of freedom are dependent on structural factors like the tolerance factor, which is generally determined by the average size of the RE/AE cations, size mismatch between the RE and AE cations and the value of x [3]. In general the reduced tolerance factor (reduced average size of RE/AE cations) reduces the itinerant $e_g$ electron bandwidth (W) and hence favours carrier localization [3–5]. Among the low W manganites, $Pr_{1-x}Ca_xMnO_3$ (PCMO) has been recognized as a prototype charge ordered/orbital ordered insulator (COI) [6–9]. The ground state of PCMO ($0.3 \leq x \leq 0.5$) is COI with a pseudo CE type antiferromagnetic (AFM) spin order. The AFM spin order transforms into ferromagnetic insulator (FMI) at $0.15 < x < 0.3$. The large size difference between the Pr/Ca and the $Mn^{3+}/Mn^{4+}$ cations leads to a small tolerance factor and transfer integral between the $Mn^{3+}$ and $Mn^{4+}$ cations. This enhances localization tendencies and stabilizes the CO/OOI phase over large range of *x*. However, at low doping ($0.15 < x < 0.3$), COI is absent and only FMI is observed. In general the ferromagnetic metal (FMM) phase is realized via COI to FMM phase transformation only under the influence of external perturbations like magnetic field [3–8]. The magnetic field required for COI-FMM phase transformation (CO melting field) is dependent on *x* and is reduced drastically as *x* is lowered from ~0.5 towards ~0.3 [9,10]. At commensurate carrier concentration (x=0.5), the COI state persists down to the lowest temperatures even at magnetic field as high as H=120 kOe but the phase boundaries shift towards lower magnetic fields at incommensurate carrier concentrations (x=0.4, 0.35 and 0.3) [7]. The neutron diffraction studies by Zirak et al. [10,11] have demonstrated that as the carrier concentration changes from commensurate to incommensurate levels the spin arrangement within the a-b plane preserves the CE-type feature but that along the c direction changes from antiparallel to parallel. The evolution of such spin order has been discussed in terms of the FM double exchange (DE) interaction along the c direction mediated by the extra electrons. Another consequence of incommensurate carrier concentration has been found to be the appearance of canted AFM at lower temperatures. In a detailed study of $Pr_{1-x}Ca_xMnO_3$ Lees et al. [12] have shown the possibility of temperature dependent multiple sources of distortion and the occurrence of magnetic field induced first order insulator-metal transition (IMT). This study also showed that the AFM ($T_N$=170 K) and CO transitions ($T_{CO}$=250 K) are decoupled around x=0.4

As outlined above, the FMM is not intrinsic to $Pr_{1-x}Ca_xMnO_3$ and external perturbations like electric and magnetic fields are needed to induce COI to FMM phase transition. However, over the last few years, it has conclusively emerged that apart from the electromagnetic perturbations, structural/microstructural modifications can also lead to the quenching of the COI and emergence of the FMM simultaneously. The possible ways to introduce structural/microstructural perturbations are RE/AE and Mn site substitutions, substrate induced strain in epitaxial thin films, downsizing of the polycrystalline material to nano-structural level and creation of oxygen vacancies.

Martinelli et al. [13] through neutron powder diffraction and magnetic measurements have shown occurrence of orthorhombic to monoclinic structural phase transition in $Pr_{0.55}Ca_{0.45}MnO_3$, which favours an AFM spin ordering at low temperature. They have also demonstrated that Cr (3 – 6 %) substitution at Mn sites can hinder this phase transition and stabilize the orthorhombic that favors FMM through DE mechanism. The role of nanostructuring enhanced surface effects on the COI has been investigated by Zhang and Dressel [14]. They have shown that at nanoscale the charge density of the itinerant $e_g$ electrons at the surface of the nanoparticles is enhanced. This drives the surface layer from COI state with collinear AFM configuration to a phase separated one consisting of charge-disordered canted AFM and FM states. The CO has been observed to completely melt as the particle size is reduce below 40 nm and has been attributed to the transformation of the core spin configuration from AFM to canted AFM through spin coupling. Rao and Bhat [15] have demonstrated that at commensurate carrier concentration (x=0.5) nanostructuring induced surface disorder suppresses the CO state, gives rise to a size-induced dominant equilibrium irreversible FM phase, provides tunability to FMM-AFM phases and induces zero field hysteretic IMT. The suppression of AFM/COI phase and simultaneous appearance of FM transition around $T_C$~105 K has been observed by Rao et al. [16] in $Pr_{0.5}Ca_{0.5}MnO_3$ nanowires. Nanostructuring induced collapse of AFM/COI phase and simultaneous occurrence of FM phase has also been observed in over doped material, e.g., $Pr_{0.4}Ca_{0.6}MnO_3$ [17]. Elovaara et al. [18] in their detailed structural and magnetic characterization of polycrystalline $Pr_{1-x}Ca_xMnO_3$ (0≤x≤0.5) have observed decrease in the unit cell volume and improvement in the structural symmetry with increase in Ca content. They have also observed the coexistence of AFM-FM spin orders at x≤0.2. At higher divalent concentrations

...

($0.2 < x \leq 0.5$) charge ordering, training effect and irreversible metamagnetic behaviour have been observed. Their results also show that the CO transition is explicit in magnetization measurements only at $x > 0.3$. Rawat et al. [19] have explored the role of change in the relative width of the supercooling/superheating band and kinetic arrest band for a FMM-AFMI transition in noncrystalline $Pr_{0.67}Ca_{0.33}MnO_3$. Their results clearly establish a correlation between the kinetic arrest band and the supercooling band has been shown experimentally, in contrast to the universal observation of anticorrelation reported so far.

The impact of substrate induced structural/microstructural perturbations on the magnetic and magnetotransport properties have been investigated by several authors [20–22]. Here we would like to mention that the strain arising owing to the distinct lattice constants of the film material and the substrate is one of the key external stimuli and has profound impact on the size and fraction of the FMM and AFM/COI phases [23–28]. The tensile strain enhances the Jahn-Teller (JT) distortion by stretching the $MnO_6$ octahedra along the film plane and hence favors the AFM/COI phase. In contrast, the compressive strain weakens the JT distortion through the elongation of the $MnO_6$ in the out-of plane direction and hence favours the FM–DE, which enhances the FMM phase [23–28]. The creation of defects due to the relaxation of strain also hinders the JT distortion and favors the FMM at the cost of AFM/COI.

De Brion et al. [20] have shown that in a 250 nm thick film of $Pr_{0.5}Ca_{0.5}MnO_3$ grown on $LaAlO_3$ substrate a segregated FM phase grows within the CO phase. This segregated phase acquires shape of very thin layers parallel to the film plane and the FM domains grow with increasing external field. Prellier et al. [21] has observed the tensile strain induced enhancement of the metastable character of the CO state in a $Pr_{0.5}Ca_{0.5}MnO_3$ thin film on $SrTiO_3$ substrate. They have also observed a hysteretic IMT at a magnetic field (H=60 kOe) much smaller than the CO melting fields required to melt the CO in polycrystalline bulk and single crystals of same composition. Okuyama et al. [22] have studied the effect of growth orientation on CO/OOI phenomena in $Pr_{0.5}Ca_{0.5}MnO_3$ epitaxial thin films and observed large enhancement in the CO transition temperature ($T_{CO} \sim 300$ K) in the film on LSAT (001), while the film on LSAT (011) showed the same $T_{CO} \sim 220$ K as the bulk.

Recently the consequences of oxygen content on the nature of coexisting magneto-electric phases have also been highlighted and it has been demonstrated that oxygen vacancies can also destabilize the COI state. It has been proposed that oxygen vacancies can introduce spin canting in the back ground to competing antiferromagnetic super exchange (AFM-SE) and FM-DE, which may devolve into a spin glass (SG) or cluster glass (CG) and add metallic flavour to the COI state. Majumdar et al. [29] have investigated the effect of oxygen on the structural and magnetic properties of low hole doped $Pr_{1-x}Ca_xMnO_3$ ($x$=0.1). Their results show that at higher oxygen content, the structural strain partially relaxes and the $Mn^{4+}$ concentration increases, thus strengthening the double-exchange interaction between $Mn^{3+}$ and $Mn^{4+}$ ions, leading to enhanced ferromagnetism. With decreasing oxygen content, also the magnetic moment increases, which could be due to the trapping of electrons at the oxygen vacant sites leading to the formation of rigid magnetic polarons. Recently Agarwal et al. [30] have shown the quenching of the CO state and appearance of hysteretic IMT in $Pr_{0.58}Ca_{0.42}MnO_3$ oriented polycrystalline thin film grown on $SrTiO_3$ (001) substrates under oxygen deficient conditions. The IMT occurs at substantial higher temperature than reported for nanostructured polycrystalline bulk samples [15]. From the above it is also clear that impact of oxygen content could be dependent on the value of $x$.

Here we would like to mention that although Agarwal et al. [30] have suggested oxygen vacancies as the cause of zero filed IMT in this robust COI system, the origin of this feature is still puzzling. Therefore further investigations are needed to clarify the origin of the CO quenching and hysteretic self-field IMT in $Pr_{0.58}Ca_{0.42}MnO_3$ polycrystalline films. In the present work we have grown two sets of oriented polycrystalline $Pr_{0.58}Ca_{0.42}MnO_3$ films on $LaAlO_3$ [LAO, (011)] substrates. One set of these films was annealed in oxygen ambient and the other set was annealed in air (under oxygen deficient condition). Our results clearly show that oxygen deficiency is conducive to the growth of FMM phase and hysteretic IMT at self-field, while oxygen annealing favours the growth of COI phase. Microstructural characterization of these films employing transmission electron microscopy unravels that the oxygen deficiency results in smaller grain size and creates localized disorders in form of lattice defects. In addition to the nano-structural effects which enhance surface effects and hence favour FMM phase, the lattice defects, which are predominant along the grain boundaries could weaken the Jahn-Teller distortion and hence favour the FMM.

**Experimental**

$Pr_{0.58}Ca_{0.42}MnO_3$ (PCMO) thin films (thickness ~300 nm) were grown by nebulized spray pyrolysis on $LaAlO_3$ (110) single crystal substrates. Stoichiometric solution of Pr, Ca, and Mn nitrates (Pr:Ca:Mn=0.58:0.42:1) in deionized water was sprayed on the substrates maintained at $T_S$~200 °C. One set of films was annealed in air and the other in flowing oxygen for a duration of 12 hr. The annealing temperature was maintained at ~1000 °C. The synthesis conditions are identical the one reported in reference 30. The gross structural characterization was done by high resolution X-ray diffraction (HRXRD, PANalytical X'Pert Pro MRD). The surface morphology was characterized by Scanning electron microscopy (SEM, Model: LEO 0440 equipped with ISIS 300 Oxford) in the secondary emission mode. The local area structural and microstructural characterization was carried out by high resolution transmission electron microscopy (HRTEM, model FEI Tecnai G2 F30 S-TWIN with the field emission electron gun sources, operated at the electron accelerating voltage of 300 kV). The cationic composition of all these films was studied by energy dispersive spectroscopy (EDS) attached to the HRTEM. The temperature and magnetic field dependent magnetization was measured by a commercial (Quantum Design) 7 Tesla SQUID magnetometer. The temperature and magnetic field dependent electrical resistivity ($\rho(H) - T$, $\rho(T) - H$) measurement was carried out in a commercial physical property measurement system (PPMS, Quantum Design). Hereafter, the air and oxygen annealed films will be referred to as PCMO-A and PCMO-O, respectively.

HRXRD data ($2\theta$-$\omega$ scan) plotted in Fig. 1 shows the occurrence of diffraction maxima alongside that of the substrate (hk0) peaks. The diffraction peaks belonging to PCMO can be indexed (assuming the orthorhombic structure [18]) either (h00) or (0k0). The out of plane lattice constants of the PCMO-A and PCMO-O are a=5.4399(1) Å and 5.4302 (6) Å, respectively. Assuming cubic symmetry, the PCMO reflections correspond to (110) and (220) planes and the corresponding out-of-plane lattice parameters of the PCMO-A and PCMO-O are a=3.847 Å and 3.840 Å, respectively. These lattice parameters are very close to that of the bulk PCMO of similar composition and hence we conclude that both the films are free of any strain. However, in few of the strong texturing local epitaxial growth and hence the presence of localized strained pockets cannot be ruled out. As revealed by the HRXRD data the oxygen annealing does not impact the gross structural features, except that it reduces the out of plane lattice constant slightly. The surface of the air annealed films (Fig. 2) mainly consists of grains of average size ≈500 nm and is found to be generally rough due to the presence of voids and outgrowths. In comparison, the average grain size in the oxygenated films is found to be larger and the surface appears denser and compact. The differences in the structural and microstructural properties of the two films could be due to reduced surface tension at the substrate-film interface during the growth in the oxygen ambient.

The temperature dependent magnetization (M-T) was measured employing zero field cooled (ZFC) and field cooled (FC) protocols at a dc magnetic field H=500 Oe applied along the longer dimension parallel to the surface of the film. The ZFC and FC M-T data of the two films are presented in Fig. 3 (a,b). In bulk form $Pr_{1-x}Ca_xMnO_3$ shows a transition to the charge ordered (CO) state around $T_{CO}$~230 K – 240 K over an appreciable range of divalent doping (x) [6–9]. However, in the present case the film prepared in air ambient (PCMO-A) does not show any explicit CO-transition. Instead a change in the slope around T~230 K is observed (inset of Fig. 3a). This suggests that in the PCMO-A film the CO has not vanished but its fraction has been reduced. The rises magnetization below $T_C^{ONSET} \approx 172$ K marks the appearance of FM correlations in the back ground of quenched CO in the PCMO-A film. The change in slope of the M-T curves around temperature marked $T_N \approx 104$ K (Neel temperature) could be attributed to the appearance of AFM correlation in the back ground of FM+COI phase. In the lower temperature region a cusp is observed at $T_P \approx 60$ K in the the ZFC curve and at $T_F \approx 40$ K magnetization shows decline. The FM transition temperature ($T_C$) was determined from the first order derivative of the ZFC M-T curve (dM/dT) and is found to be $T_C \approx 148$ K. From the data presented in Fig. 3a it is clear that the appearance of FM correlations is accompanied by a huge divergence between the ZFC-FC branches. Such bifurcation of the magnetizations curves is generally regarded as the signature of the cluster glass like metamagnetic state [30–33] and has origin in the coexistence of different magnetic phases like FMM and AFM/COI.

In contrast to the PCMO-A, the film prepared in flowing oxygen shows a robust signature of CO at $T_{CO} \approx 236$ K (Fig. 3b). The CO phase is followed by appearance of FM correlations at $T_C^{ONSET} \approx 172$ K and the FM transition temperature as determined from the dM/dT curve is $T_C \approx 148$ K. The AFM correlations appear at $T_N \approx 140$ K. The ZFC curve shows a cusp at $T_P \approx 68$ K and at $T_F \approx 48$ K the magnetization shows decline. The ZFC-FC divergence in this film is also similar to the PCMO-A film. The major difference between the magnetization behaviour of the two films grown under different ambient is that the air annealed film has significantly larger magnetic moment than the oxygenated one. Interestingly, despite much lower magnetic moment the $T_C$ of the PCMO-O film is higher than its air annealed counterpart. Another difference is the significantly higher value of the Neel temperature ($T_N$) in the PCMO-O film. This clearly shows that the PCMO-A film has much higher FM phase fraction than the PCMO-O film. This is well corroborated by the quenching of the CO phase in the PCMO-A film. From the data presented above it is clear that growth in oxygen ambient favours the AFM/COI phase, while air ambient (growth under reduced oxygen concentration) is conducive to the FMM phase. The magnetization behaviour of the two films was also probed by isothermal magnetization-magnetic field (M-H) measurements. The M-H loop of the two films is plotted in Fig. 4. In the initial (virgin) cycle the M-H curves of the PCMO-A and PCMO-O films show a sharp rise in magnetic moment around H≈25 kOe and 35 kOe, respectively. The virgin cycle M-H curve of these films do not show any saturation upto H=70 kOe. The M-H curves do show saturation in subsequent cycles and a well-defined M-H loop is also observed. The saturation magnetic moment of the PCMO-A and PCMO-O films are found to be $M_s \approx 9$ $\mu_B$ and 8 $\mu_B$, respectively. These values are much higher than the theoretical value of the magnetic moment. We believe that such abnormally high magnetic moment could be regarded as the signature of the giant magnetic moment effect [31]. The giant moment effects have been reported for several magnetic glass forming alloys and has been attributed to the field induced magnetic polarization effect, wherein the non-FM phases like AFM are transformed in FM ones [31]. In the present case, the jump in the magnetic moment clearly suggests the presence of a field induced magnetic polarization effects in which the AFM/COI phases could be transformed into FM ones.

The temperature dependent of resistivity ($\rho$ - T) was measured in the temperature range 300 K – 4 K – 300 K in cooling as well as warming cycles at H = 0 kOe, 10 kOe, 30 kOe and 50 kOe. The $\rho$ - T of PCMO-A is plotted in Fig. 5. At room temperature the zero field resistivity is $\rho$ (H=0) =0.015 $\Omega$-cm. As the temperature is lowered the resistivity increases and the $\rho$ - T curve changes slope around 230 K but the CO ordering transition, which is generally observed around $T_{CO}$ ~ 230-240 K [6–9], is not explicit. In the cooling cycle insulator-metal transition (IMT) occurs at $T_{IM}^C$ (H=0)≈89 K. At T=4.2 K the resistivity ($\rho$~0.3 $\Omega$-cm) is more than an order of magnitude higher than that at room temperature and this is generally regarded as a signature of the percolative nature of the electrical transport in manganites [8,34]. In warming cycle the $\rho$ - T traces a different path and the IMT occurs at significantly higher temperature $T_{IM}^W$ (H=0)≈148 K. This results in strong thermal hysteresis between cooling and warming cycle $\rho$ - T curves. The $\rho$ - T behaviour of this films is nearly similar to the PCMO film of same composition prepared on SrTiO$_3$ (001) substrates [30]. In external magnetic field (i) resistivity is dramatically reduced, (ii) both cooling and warming cycle IMTs ($T_{IM}^C$ and $T_{IM}^W$) increases, (iii) difference between the $T_{IM}^C$ and $T_{IM}^W$ is gradually reduced, (iv) thermal hysteresis in the $\rho$ - T is reduced, and (v) at sufficiently high value of magnetic field $T_{IM}^C$ and $T_{IM}^W$ coincide. For example, $T_{IM}^C$ and $T_{IM}^W$ are observed to increase to ~112 K and ~162 K at H= 10 kOe and coincide at H=50 kOe to $T_{IM}$~216 K. The variation of $T_{IM}^C$ and $T_{IM}^W$ is shown in the inset of Fig. 5. The prominent hysteresis in the $\rho$ - T curve is a signature of strongly phase separated state, which in view of the magnetic properties elucidated earlier, is composed of the AFM/COI and FMM phases. The gradual decrease in the $\rho$ - T hysteresis with applied magnetic field clearly supports the above conclusion and is a signature of the decrease in the degree of phase separation due to the field induced AFM/COI to FMM phase transition.

The $\rho$ - T measurement of the film annealed in oxygen (PCMO-O) reveals several distinct features (Fig. 6). In absence of magnetic field the room temperature resistivity is $\rho \approx 0.32$ $\Omega$-cm, which is an order of magnitude higher than that of the PCMO-A film. The slope of the $\rho$ - T curve changes sharply in the vicinity of the $T_{CO}$ and is followed by a broad hump around T~200 K. The zero filed resistivity becomes immeasurable below T~89 K. At T<200 K the cooling and warming cycle $\rho$ - T shows irreversible behaviour, which could be due to the presence of small FMM clusters in the AFM/COI matrix. No noticeable change is observed in the $\rho$ - T curve at H=10 kOe. As magnetic field is increased to H=30 kOe, an IMT appears at $T^{C}_{IM} \approx 66$ K and $T^{W}_{IM} \approx 144$ K in the cooling and warming cycles, respectively. The $\rho$ - T data taken at H=30 kOe shows strong thermal hysteresis with cooling and warming cycle curves showing resistivity jumps. These $\rho$ - T discontinuities are more prominent in the cooling cycle curve and could be attributed to the presence of a metamagnetic state caused by the magnetic frustration resulting from the coexisting FMM and AFM/COI phases. The prominence of the resistivity jumps in the cooling cycle also shows that during the cooling cycle the magnetic frustration owing to the coexisting AFM/COI and FMM phases is much stronger than that in the warming cycle. This could be attributed to the difference in the kinetics of the H induced AFM/COI to FMM transformation during the cooling cycle and the reverse transformation during the warming cycle. As evidenced by the magnetization data, the temperature regime in which these $\rho$ - T discontinuities occur is akin to cluster glass. At H=50 kOe, IMT shows large enhancement, being $T^{C}_{IM} \approx 169$ K and $T^{W}_{IM} \approx 190$ K in the cooling and warming cycles, respectively. It is interesting to note that the thermal hysteresis in the $\rho$ - T curve remains very prominent even at H = 50 kOe. Thus it is clear that the temperature dependent magnetotransport properties of the PCMO-O film differ drastically from that of the PCMO-A film. The observed differences could be understood in terms of the difference in the magnetic phase profile of the two films. As demonstrated by the magnetization data presented earlier, the PCMO-A film has higher (lower) fraction of the FMM (AFM/COI) phase, while the dominant magnetic phase in the PCMO-O film is AFM/COI. The dominance of the AFM/COI phase over the FMM in PCMO-O film leads to (i) appearance of IMT at H$\geq$30 kOe with much stronger thermoresistive hysteresis, (ii) occurrence of jumps in the ρ - T curve at intermediate magnetic fields and (iii) non equality of $T^C_{IM}$ and $T^W_{IM}$ even at H=50 kOe.

In order to further elaborate the electrical transport in these films we measured the magnetic field dependent resistivity (ρ - H) in the range H = ±60 kOe at several temperatures; ranging from 5 K to 200 K. The ρ - H data of the PCMO-A film are plotted in Fig. 7 and 8. First we describe the evolution of the ρ - H data measured at T=5 K. In the virgin cycle, ρ (H=0)≈0.03 Ω-cm and it first decreases slowly as H is increased and approaches ≈0.024 Ω-cm at H=25 kOe. At H>25 kOe, ρ (H) decreases by more than an order of magnitude to 0.0018 Ω-cm at H=49 kOe. Beyond this value of H, ρ (H) decreases gradually to ≈0.0016 Ω-cm. As H is traced back to zero, the virgin cycle ρ - H curve is not recovered. The increase and decrease in ρ (H) with decreasing and increasing H in subsequent cycles is marginal only. It is also interesting to note that the ρ - H curve measured at T=5 K does not show any hysteretic behaviour. The M-H (the virgin cycle M-H is also not retraceable) and M-T data presented earlier clearly shows that the low temperature state of this film is akin to a metamagnetic cluster glass. The behaviour of the virgin cycle M-H and ρ-H data clearly suggest towards the frozen nature of the magnetic clusters and the transformation of the metamagnetic cluster glass to a crystalline FMM phase by the applied magnetic field; this transformation occurs in the range 25 kOe to 49 kOe. The absence of the hysteresis loop could be attributed to temporal relaxation effects.

With rise in temperature softening of the frozen cluster glass is expected. This feature is unambiguously demonstrated by the ρ - H curve taken at T=40 K. As seen in Fig. the virgin cycle ρ (H) decrease slowly from 0.042 Ω-cm to 0.041 Ω-cm as H is increase from 0 to 3 kOe and then decreases sharply, by about two orders of magnitude to 0.0006 Ω-cm at H=60 kOe. In the subsequent H cycle the original zero file resistivity is never recovered but the ρ - H curve shows a prominent hysteresis. In contrast to the virgin cycle ρ - H data measured at T=5 K, the drop in ρ (H) in the ρ - H curve measured at T= 40 K is sharper and occurs at very small value of H. This clearly shows that the cluster glass like state is softened by the increased thermal energy. The hysteresis in the subsequent cycles is caused by the difference in the H induced conversion of the AFM/COI phase to FMM in the H increasing cycle and reverse transformation (FMM-AFM/COI) during the H decreasing cycle. As shown in Fig. 7, during the H decreasing cycle, ρ (H) increases very slowly till H~34 kOe and then it grows faster. This suggests that as H is lowered from 60 kOe, the FMM to AFM/COI transition is blocked till H~34 kOe and then unblocked. Similarly, during the field increasing cycle (other than the virgin cycle), the decrease in ρ (H) is rather slow till H~30 kOe and beyond that it becomes slightly faster. This clearly shows that AFM/COI to FMM conversion is blocked till H~30 kOe.

At higher temperatures, the ρ - H curves still show prominent hysteresis and the drop in resistivity as a function of H is observed to increase gradually with increase in temperature. The sharpest change in ρ - H is observed in the vicinity of the $T_{IM}$. This is clearly evidenced by the ρ - H curves measured at 120 K and 150 K (Fig. 7), which show more than two orders of magnitude decrement in the resistivity at H=60 kOe. As the temperature is increased further, e.g., T=200 K, the resistivity decreases slowly till H≈20 kOe and then drops rapidly; the overall behaviour remaining nearly non-hysteretic. This could be attributed to the H induced transformation of COI clusters into FMM ones. This ρ - H behaviour, however, is different from that observed in the lower temperature regime, e.g., at T= 5 K and 40 K. The lower temperature region is a dominantly phase separated cluster glass. In contrast, as the temperature is increased the density of FMM clusters which are believed to appear well above the above $T_C/T_{IM}$ decreases far below the threshold level required to create a strong phase separated state. Hence the temperature regime well above $T_C/T_{IM}$ is dominantly a charge order one and the same is well explained by the magnetization data presented earlier. Under the influence of a magnetic field the AFM/COI clusters are transformed into FMM in the H increasing cycle and in the reverse process, the FMM clusters acquire their original, that is, AFM/COI configuration. Both these process are linear and hence nonhysteretic.

The ρ - H data of the PCMO-O film taken at different temperature are plotted in Fig. 9 and 10. At T=5 K, the general behaviour of the ρ - H curve is nearly the same as observed for the PCMO-A film. But the major difference is seen in the behaviour of the virgin cycle ρ - H curve. As seen in Fig. 8, at H=0 the resistivity was extremely high and could not be measured. Even at H=36 kOe, the resistivity is as large as ρ≈1.4 kΩ-cm and it decreases slowly till H=41 kOe. At this point ρ decreases from 910 Ω-cm at H=41 kOe to 0.01 Ω-cm at H=49 kOe; a decrease of about five orders of magnitude within a very narrow range of magnetic field. At H>49 kOe, the resistivity increases rather slowly and appears to saturate at H=60 kOe. As H is traced back to zero, the virgin cycle ρ - H curve is not recovered. The increase and decrease in ρ (H) with decreasing and increasing H in subsequent cycles is small but larger than that observed for the PCMO-A film. Like the PCMO-A film the ρ - H curve measured at T=5 K does not show any hysteretic behaviour. The much larger value of the magnetic field required to melt the cluster glass state into a FMM one clearly suggest that in the oxygen annealed film the cluster glass is much more stronger than that in the air annealed film. As the temperature is increased, the resistivity decreases but still remains immeasurable in zero magnetic field. At T=40 K, ρ (H) after initial slow decrease, drops by about three orders of magnitude within a very narrow H window, e.g., from 360 Ω-cm at H=27 kOe to 0.35 Ω-cm at H=35 kOe. At this temperature too, the virgin cycle ρ (H) is not retraceable in subsequently H cycles. In H increasing cycles (other than the initial cycle), the initial decrease in ρ (H) is rather slow till H=21 kOe and at higher fields the decline becomes faster; the net decrease in being more than an order of magnitude at H=60 kOe. During H decreasing cycle, ρ (H) varies very little till H=27 kOe and then increases rather rapidly. This shows that the FMM clusters are blocked/pinned till H=27 kOe in the H decreasing cycle and hence the FMM – AFM/COI conversion being slow. Similarly, the AFM/COI – FMM conversion is also blocked till H=21 kOe during the H increasing cycle.

In contrast to the air annealed film, ρ - H data of the PCMO-O taken at higher temperatures (Fig. 10) shows several interesting features which provide evidence in favour of a metamagnetic state in it. The ρ - H curve measured at T=80 K, in addition to the large hysteresis, shows several abrupt jumps in the H increasing as well as decreasing cycle. In the H increasing cycle, ρ (H) is observed to decrease slowly till H= 28 kOe and then declines from ≈112 Ω-cm at H=35 kOe to 1.4 Ω-cm at 39 kOe. At H > 39 kOe the decrease is representative of a prototype FMM manganite. Here we would like to mention that the measurements were performed several times and the discontinuities and jumps in ρ - H curves have been found to be reproducible barring very small shifts in the values of ρ and H. The ρ - H curve measured at T=100 K also shows metamagnetic jumps but the discontinuities and jumps in the ρ - H curve are not as prominent as in that measured at 80 K. But the hysteresis and the decrements in the resistivity are nearly of the same order. As the temperature increases to further higher values, the ρ(H) decrement as well as the hysteresis decreases occurs (e.g. the ρ -H curves measured at 120 K and 150 K). At further higher temperature, e.g., T=200 K, the resistivity decreases slowly till H≈20 kOe and then drops rapidly; overall behaviour remaining non-hysteretic. As point out earlier, this ρ - H behaviour, however, is different from that observed in the lower temperature regime, e.g., at T= 5 K and 40 K and could be attributed to the H induced transformation of COI clusters into FMM ones. From the ρ-T and ρ-H data it is clear that the major difference between the two sets of films (PCMO-A and PCMO-O) is the prominence of the metamagnetic component in the PCMO-O film. This could be attributed to the higher degree of magnetic frustration caused by larger fraction of the AFM/COI in this film.

In order to differentiate the nature of the phase separated state in the PCMO-A and PCMO-O film, we have measured the area of the $\rho$ - H loops taken at different temperatures and analyzed its temperature dependence. For the calculation of the loop area, the magnetic field dependent resistivity [$\rho$ (H)] was normalized by the value at H=0 [$\rho$ (H=0)] and then $\frac{\rho(H)}{\rho(0)}$ was plotted as a function of the magnetic field. The temperature dependence of the normalized loop area is plotted in Fig. 11. In the lower temperature region, the loop area of the PCMO-A film is significantly lower than that of the PCMO-O. The difference between the loop area of these films decreases with increase in temperature and at T=200 K the loop area of both the films lie very close. The higher value of the loop area of the PCMO-O film could be regarded as a consequence of the higher degree of magnetic frustration due to the higher fraction of the AFM/COI clusters.

As discussed earlier the FMM phase is not intrinsic to $Pr_{1-x}Ca_xMnO_3$ (0.3≤x≤0.5) and that it is induced by microstructural modifications like nanostructuring, substrate induced strain, oxygen vacancies, etc. This suggests that the observed differences in the magnetic and magnetotransport properties of the PCMO-A and PCMO-O films in the present case could be due to the difference in microstructural features. Since the present films are highly oriented but polycrystalline, the presence of long range strain is ruled out. However, the localized strained regions, which in case of LAO substrate are expected to be of compressive nature, could act as inhomogeneity and cause quenched disorder. Oxygen vacancies are known to destabilize the COI phase both in single crystalline as well as polycrystalline materials quite efficiently [30,32,35]. It has been suggested that since the oxygen stoichiometry is related to the effective hole concentration therefore even a mild inhomogeneity/gradient in them could result in spatially varying carrier density that may also act as quenched disorder. Thus the strain inhomogeneity and coupled with oxygen vacancies created due to the high temperature annealing in oxygen deficient ambient could destabilize the COI state. This has been already demonstrated by Srivastava et al. [32] in oriented $Sm_{1-x}Sr_xMnO_3$ (x~0.45) thin films. We also believe that oxygen deficient ambient could give rise to different type of microstructure and that could in turn be conducive to the growth of FMM correlations. However, there is no direct evidence for change in the microstructural features as the annealing ambient is changed, e.g., from air to oxygen. To elucidate the microstructural contrast between the two sets of films, we carried out the HRTEM investigation of these two set of films.

Figure 12(a-c) exhibits a set of micrographs of PCMO-O, delineating a uniform feature with fine grains throughout dispersed in the microstructure. The nano-grained microstructure is random with clear distinctions of grain boundaries separated with boundaries often consisted of triple junctions (Fig. 12a). A clarity of fine grained microstructure is further discerned in Fig. 12b, by recording the microstructure at higher magnification to reveal further on grains and associated boundaries. At atomic scale we have observed stacking of planes with random orientation depending on the directions of individual grain growth during synthesis (Fig. 12a-c). Although as an illustrative example a set of planes with interplaner spacing of 0.27 nm [(hkl):(200)/(020)], having orthorhombic crystal structure [lattice constants: a ≈ 0.542 nm, b ≈ 0.543 nm and c ≈ 0.765 nm, ref. no. 18] with an orientation of about 72° between them is displayed as in Fig. 12c. A set of moiré fringes depicted as evolved due to overlap of few such tiny crystals at the interface are encircled by a white dotted line (Fig. 12c). It is important to mention that confining few grains in the reciprocal space, the selected area electron diffraction pattern results a pattern consisted of selective Debye rings, however individual rings is normally constituted of spots delineating that the individual grains are reasonably large in size in spite of their nano-scaled dimensions (Fig. 12d). However by considering a large number of grains while recording selected area electron diffraction pattern, sharp rings corresponding to several crystallographic planes are reflected in reciprocal space (inset in Fig. 12d). A set of crystallographic planes with Miller indices (hkl): (002), (022), (040), and (410), respectively are marked on Debye rings (Fig. 12d). There are certain subtle features different than in case of the PCMO-A. Figure 13 (a-d) delineates the main microscopic results of PCMO-A film obtained under HRTEM. It was noted that under air atmosphere, the microstructure of the film was little mushy, in the sense that instead of simple grains separated by the boundaries, there are certain types of imperfections leading to grain refinement is obvious (Fig. 13a-b). The grain size normally varies between 30 to 50 nm with lots of perturbations on the grain boundaries (Fig. 13b). At high magnifications, it is possible to observe that the microstructure is full of moiré fringes and several other kinds of imperfections (regions marked as A and B in Fig. 13c). Moreover we have interestingly observed thick fringes in the microstructure (marked as B in Fig. 13c). These thick fringes are presumably evolved due to the magnetic anisotropy present in the microstructure. The region A is further magnified to reveal the two sets of planes, (hkl): (020)

and (040), corresponding to interplaner spacing of 0.27 and 0.14 nm, respectively (inset in Fig. 13c). A corresponding selected area electron diffraction pattern recorded in reciprocal space demonstrates that the microstructure is ultrafine grained with random orientation of individual grains. A few Debye rings in the electron diffraction pattern are indexed and marked as (hkl): (020), (022), (040), (410) and (006), respectively (Fig. 13d). The sharp and continuous Debye rings in Fig. 13d further reveals that the corresponding microstructure is finer than the microstructure as obtained in case of PCMO-O film.

From the above results it is clear that annealing ambient has a direct bearing of the microstructural features of the PCMO films. Annealing in oxygen ambient results in microstructures with much less defect and disorder, while annealing in air results significantly higher density of the lattice defects and Moire fringes and the more disordered nature of the grain boundaries. Here we would like to mention that the Moire patterns are related to formation of interfaces between misaligned crystals. This could happen when thin layers parallel to the film plane are present and misaligned. In fact this scenario is very close the one proposed by De Brion et al. [20] in which the segregated FM phase grows in form of very thin layers parallel to the film plane. The presence of higher degree of disorder and defects in the air annealed film (PCMO-A) coupled with the oxygen vacancies and strain inhomogeneity could weaken the Jahn-Teller distortion, which invariably favours AFM spin order through the superexchange mechanism, and enhance the FM correlations.

**Conclusions**

The impact of growth conditions on microstructural, magnetic and magnetotransport properties of $Pr_{1-x}Ca_xMnO_3$ ($x\approx0.42$) films have been studied. Our results clearly show that the electronic phases such as AFM and AFM/COI, their relative fractions and the associated transitions are extremely sensitive to the growth conditions. Oxygen deficient growth conditions lead to quenching of the COI and favours the FMM phase. In contrast, oxygenation favour the AFM/COI and weakens FMM. Consequently films grown under oxygen ambience do not show self-field IMT. The excess of AFM/COI phase fraction in the oxygen annealed film enhances the degree of freezing of the cluster glass in the lower temperature region and leads to metamagnetic jumps and hysteretic behaviour in the temperature and magnetic field dependent resistivity. This is confirmed by the suppression of metamagnetic jumps and resistivity hysteresis at higher magnetic fields. The origin of the FMM phase high enough to induce self-field IMT in this material is traced to the microstructural modification such as finer grain size, higher fraction of Moiré fingers, etc. caused by the oxygen deficient growth condition.

**Acknowledgements**

Authors are grateful to Prof. R. C. Budhani for his persistent encouragement. Financial support from CSIR is thankfully acknowledged. VA is grateful to CSIR for the award of a senior research fellowship.

**Figure captions**

Figure 1.   High resolution X-ray diffraction pattern (2θ-ω scan) of $Pr_{0.58}Ca_{0.42}MnO_3$ films annealed in air and oxygen ambient.

Figure 2.   SEM pictures showing the surface morphology of $Pr_{0.58}Ca_{0.42}MnO_3$ films (a) annealed in air and (b) annealed in oxygen.

Figure 3.   Temperature dependent Zero Field Cooled (ZFC) and Field Cooled (FC) magnetization of $Pr_{0.58}Ca_{0.42}MnO_3$ films (a) annealed in air and (b) annealed in oxygen. The inset in figure a shows the $M^{-1}$-T graph of the film.

Figure 4.   M-H loop of $Pr_{0.58}Ca_{0.42}MnO_3$ films taken at T=10K.

Figure 5.   Temperature dependent resistivity of air annealed $Pr_{0.58}Ca_{0.42}MnO_3$ film measured in heating and cooling cycles at different magnetic fields (H=0, 10, 30 and 50 kOe). The inset shows the variation of IMT with H measured in heating and cooling cycles with increasing magnetic field.

Figure 6.   Temperature dependent resistivity of oxygen annealed $Pr_{0.58}Ca_{0.42}MnO_3$ film measured in heating and cooling cycles at different magnetic fields (H=0, 10, 30 and 50 kOe).

Figure 7.   Magnetic field dependent resistivity of air annealed $Pr_{0.58}Ca_{0.42}MnO_3$ film measured in five cycles in the range -60 kOe≤H≤60 kOe at T = 5 K and 40 K.

Figure 8.   Magnetic field dependent resistivity of air annealed $Pr_{0.58}Ca_{0.42}MnO_3$ film measured in five cycles in the range -60 kOe≤H≤60 kOe at T = 80 K, 100 K, 120 K, 150 K and 200 K.

Figure 9.   Magnetic field dependent resistivity of oxygen annealed $Pr_{0.58}Ca_{0.42}MnO_3$ film measured in five cycles in the range -60 kOe≤H≤60 kOe at T = 5 K and 40 K.

Figure 10.   Magnetic field dependent resistivity of $Pr_{0.58}Ca_{0.42}MnO_3$ film grown in oxygen ambient measured in five cycles in the range -60 kOe≤H≤60 kOe at T = 80 K, 100 K, 120 K, 150 K and 200 K.

Figure 11.   Temperature dependence of the normalized ρ-H loop area of $Pr_{0.58}Ca_{0.42}MnO_3$ films.

Figure 12.   HRTEM images of $Pr_{0.58}Ca_{0.42}MnO_3$ film grown in air ambient. Images (a) and (b) depict the microstructure taken at different magnifications. Image (c)

shows the direct lattice image. The encircle regions shows the Moiré fringes. Image (d) shows the selected are electron diffraction pattern and the inset shows electron diffraction pattern covering several grains. (Details are discussed in the text)

Figure 13. HRTEM images of $Pr_{0.58}Ca_{0.42}MnO_3$ film prepared in oxygen ambient. Images (a) and (b) depict the microstructure taken at different magnifications. High density of Moiré patterns is clearly seen in image (b). Image (c) shows the direct lattice image consisting of different lattice spacing. The thicker fringes marked 'B' represent lattice imperfections different than the Moiré fringes. Image (d) shows the electron diffraction pattern covering several grains. (Details are discussed in the text)

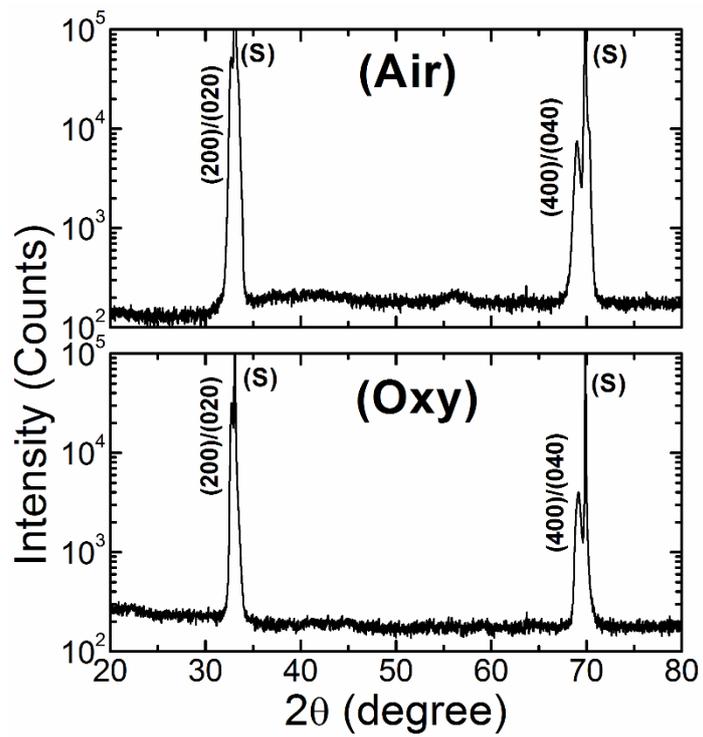

Fig. 1

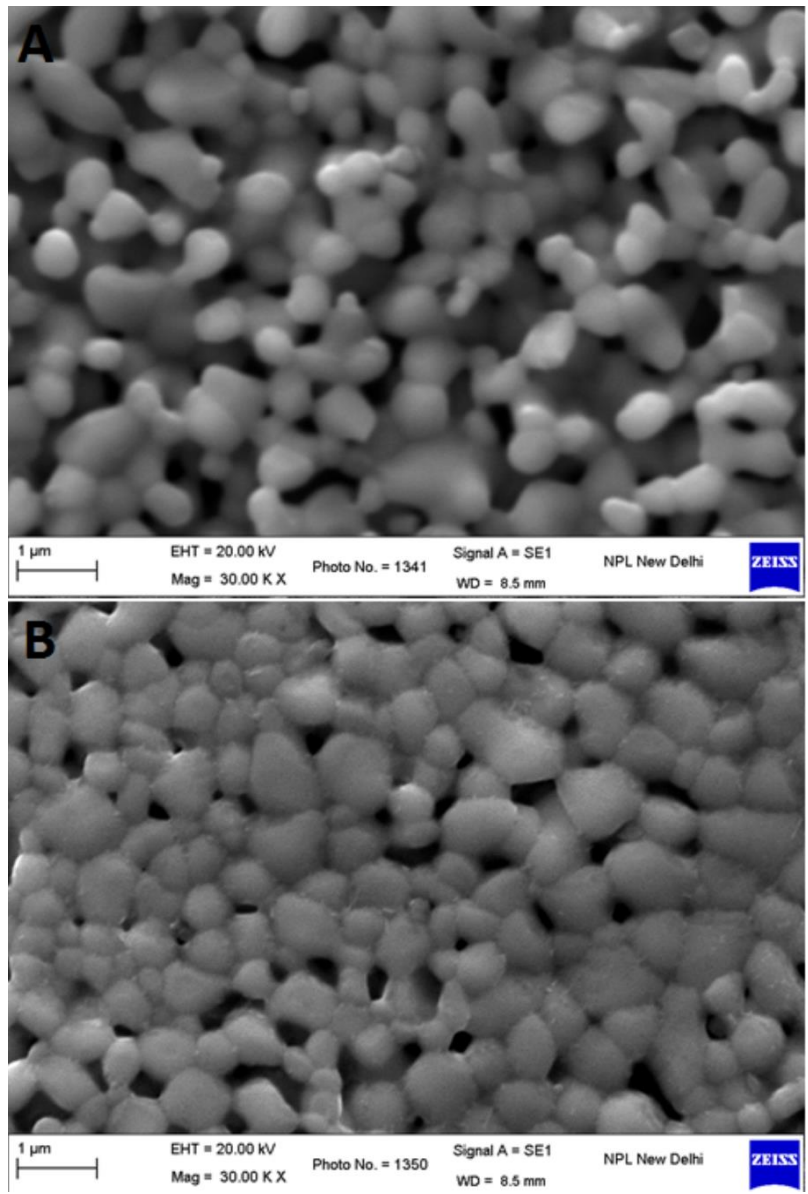

Fig. 2

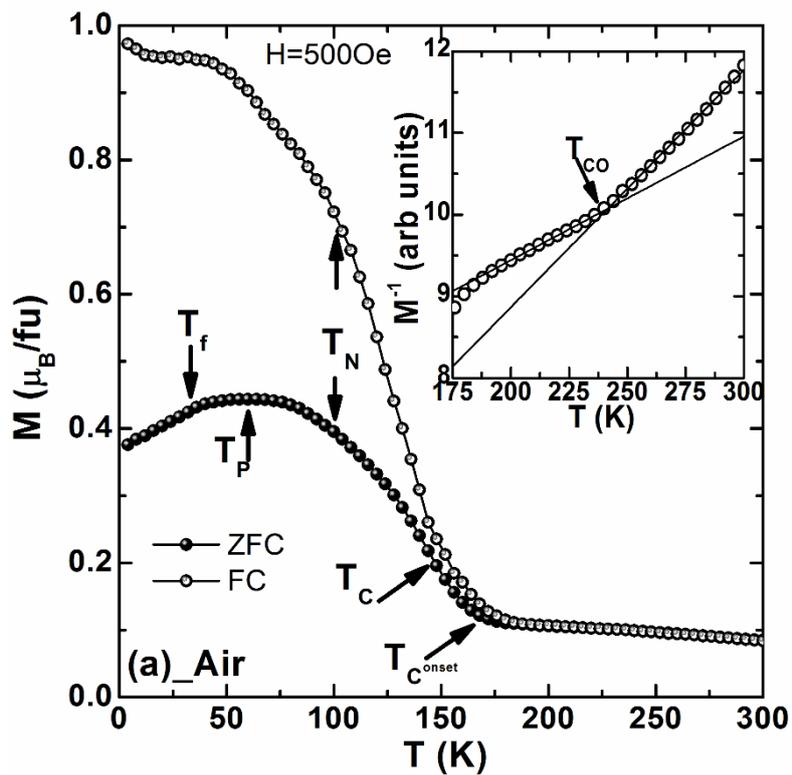

Fig. 3a

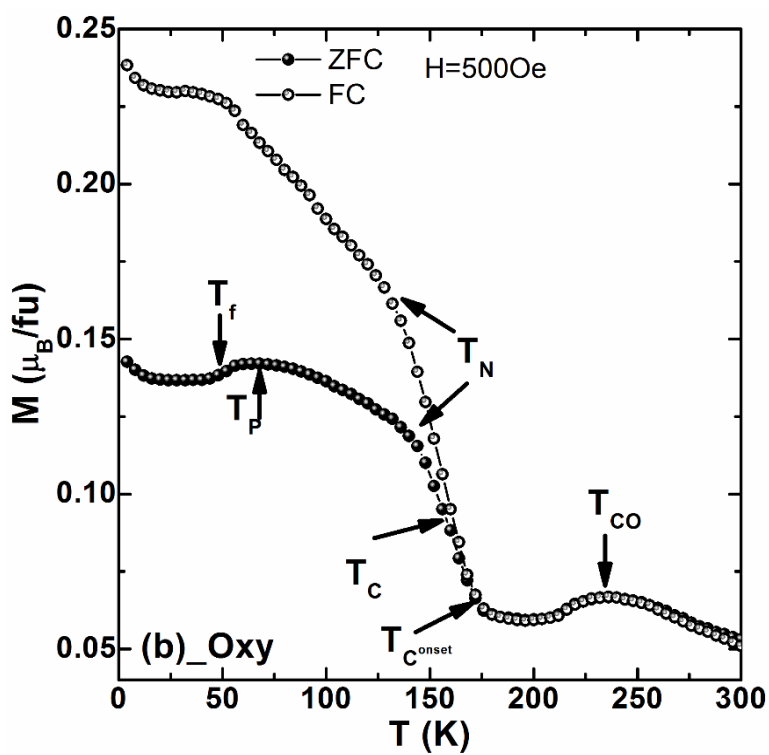

Fig. 3b

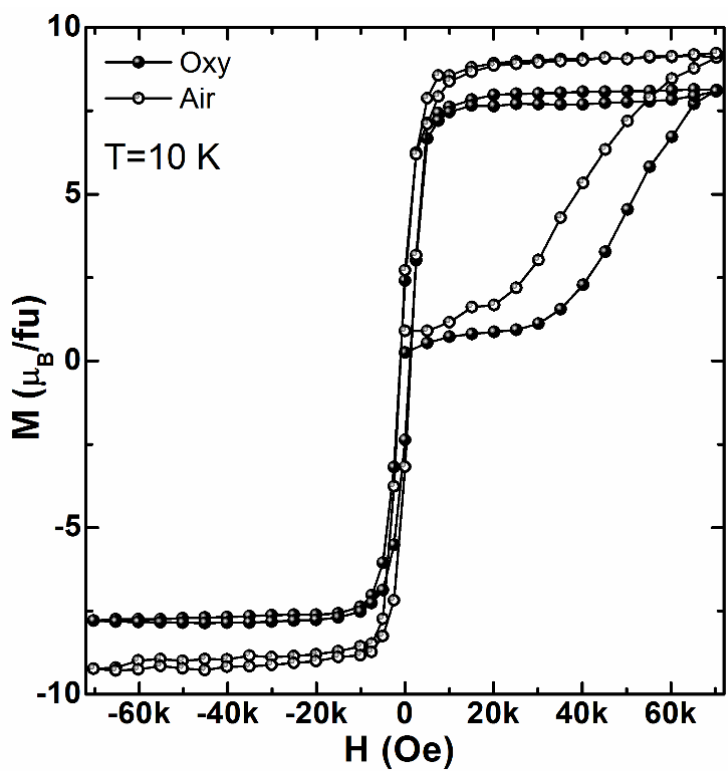

Fig. 4

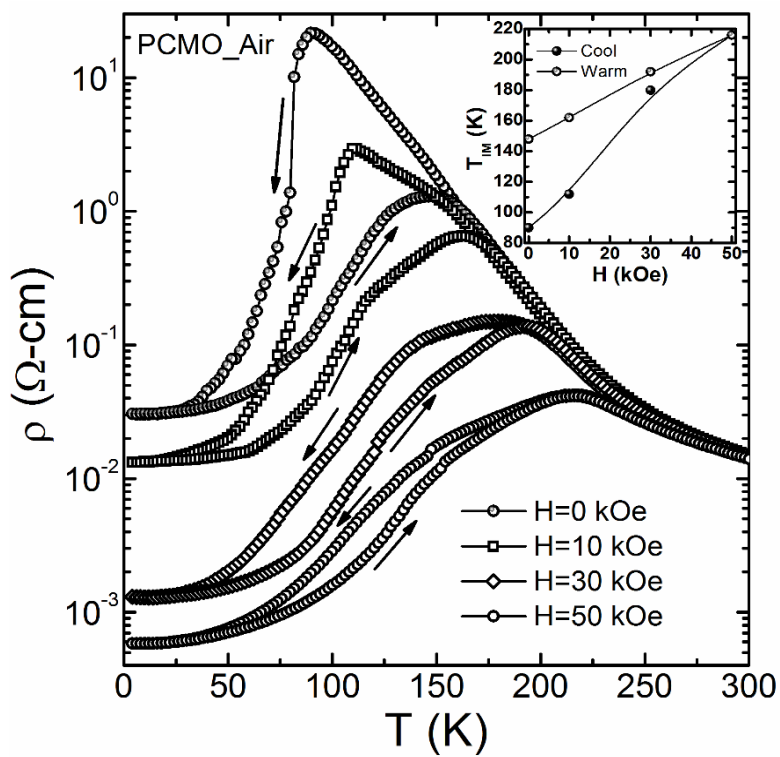

Fig. 5

24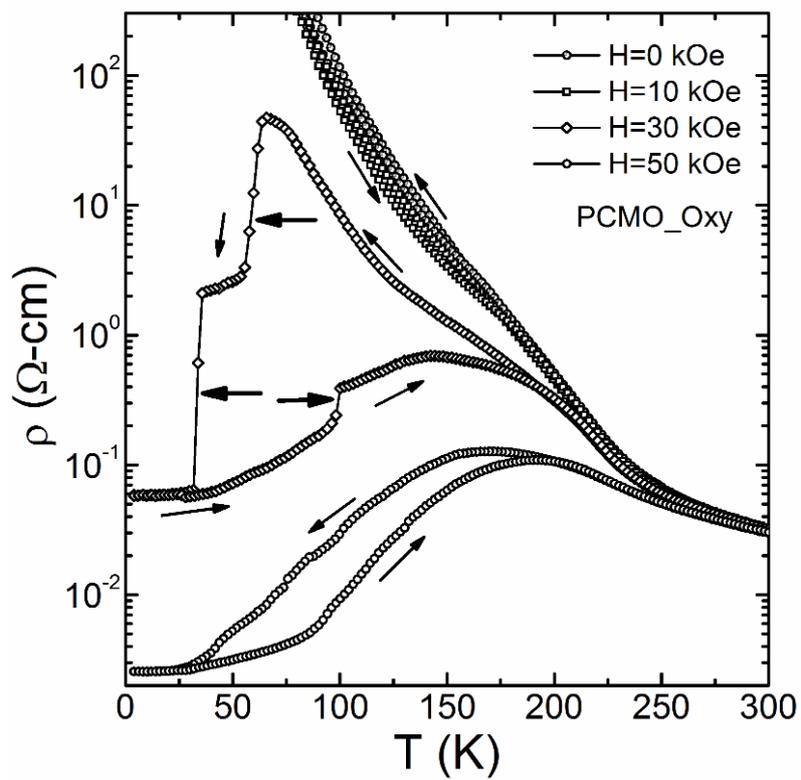

Fig. 6

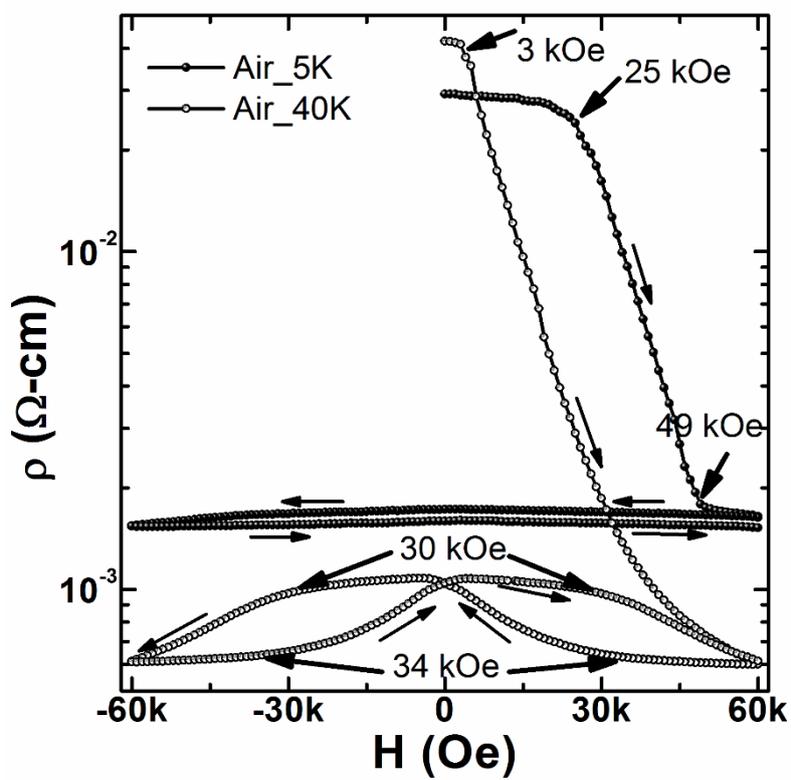

Fig. 7

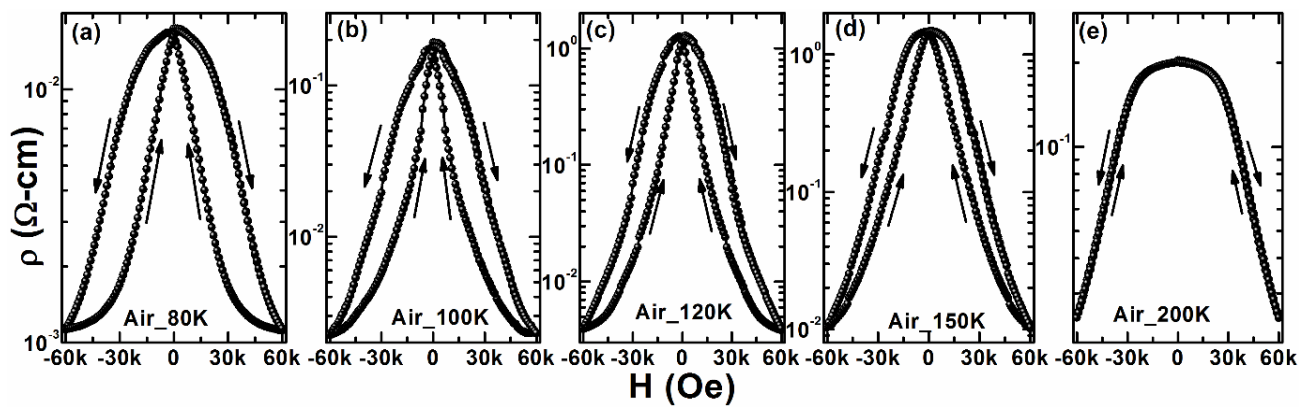

Fig. 8

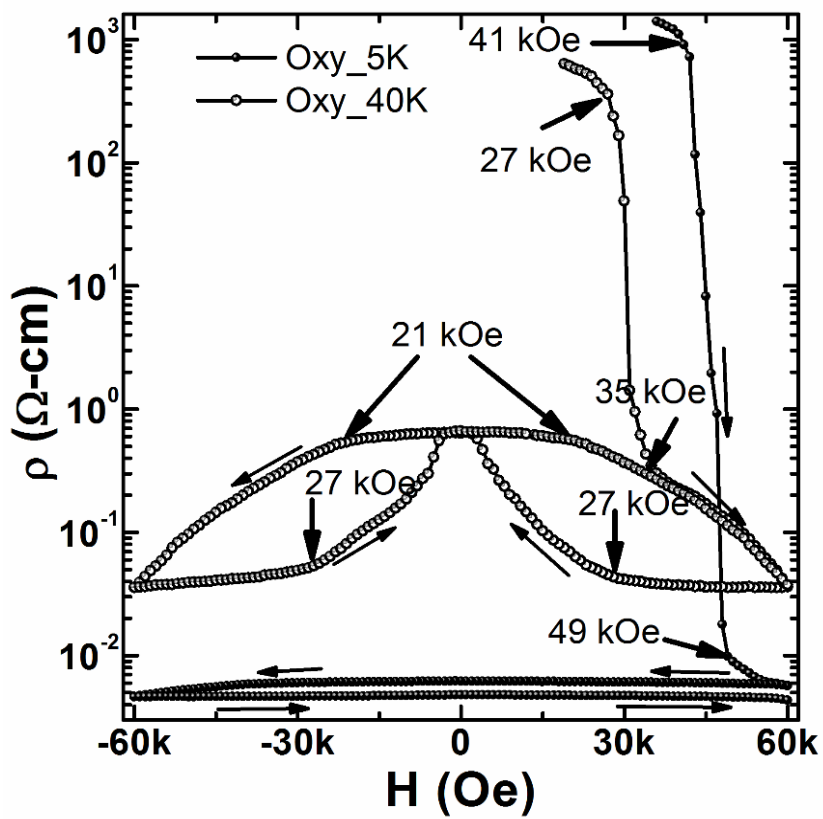

Fig. 9

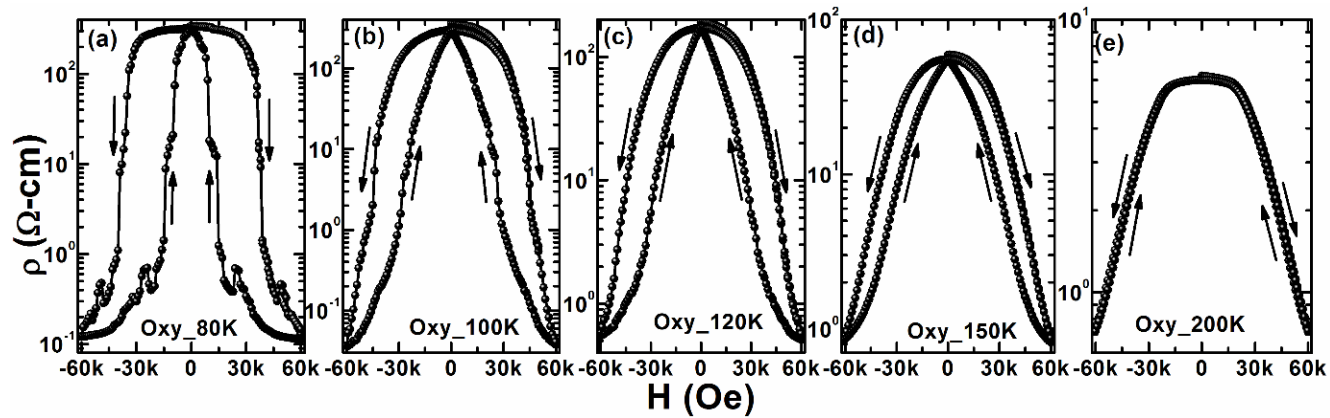

Fig. 10

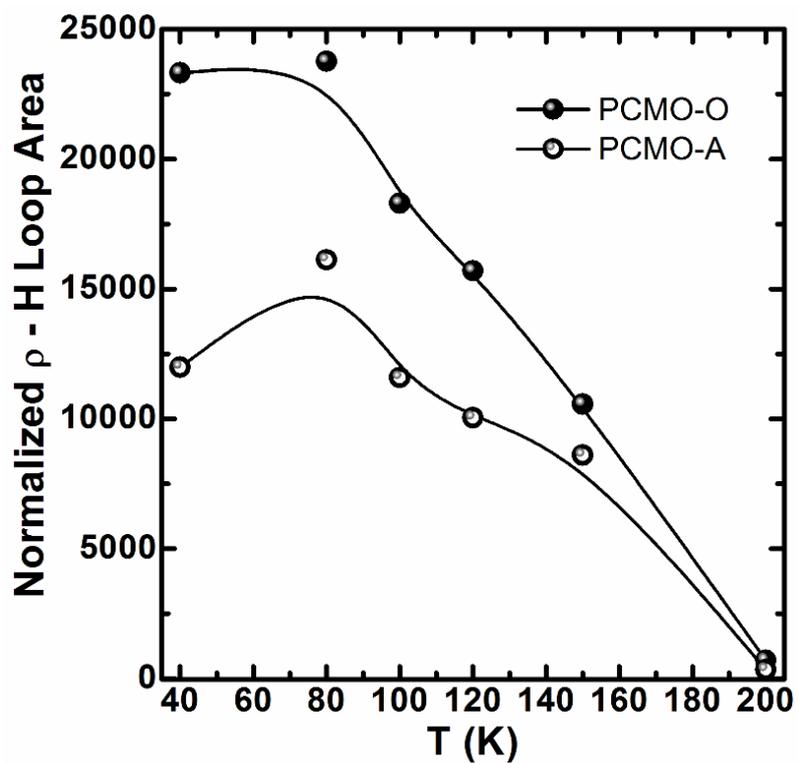

Fig. 11

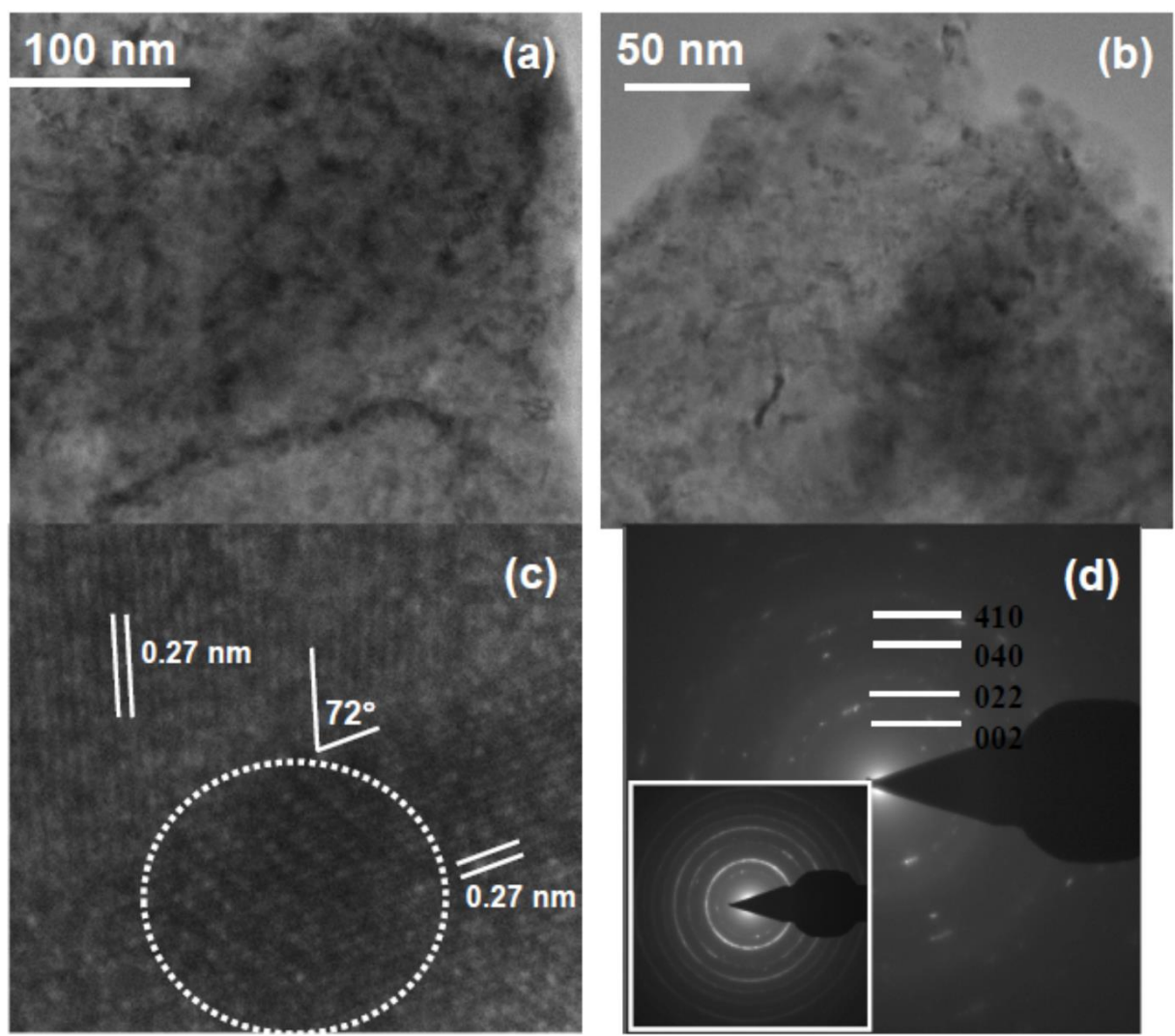

Fig. 12

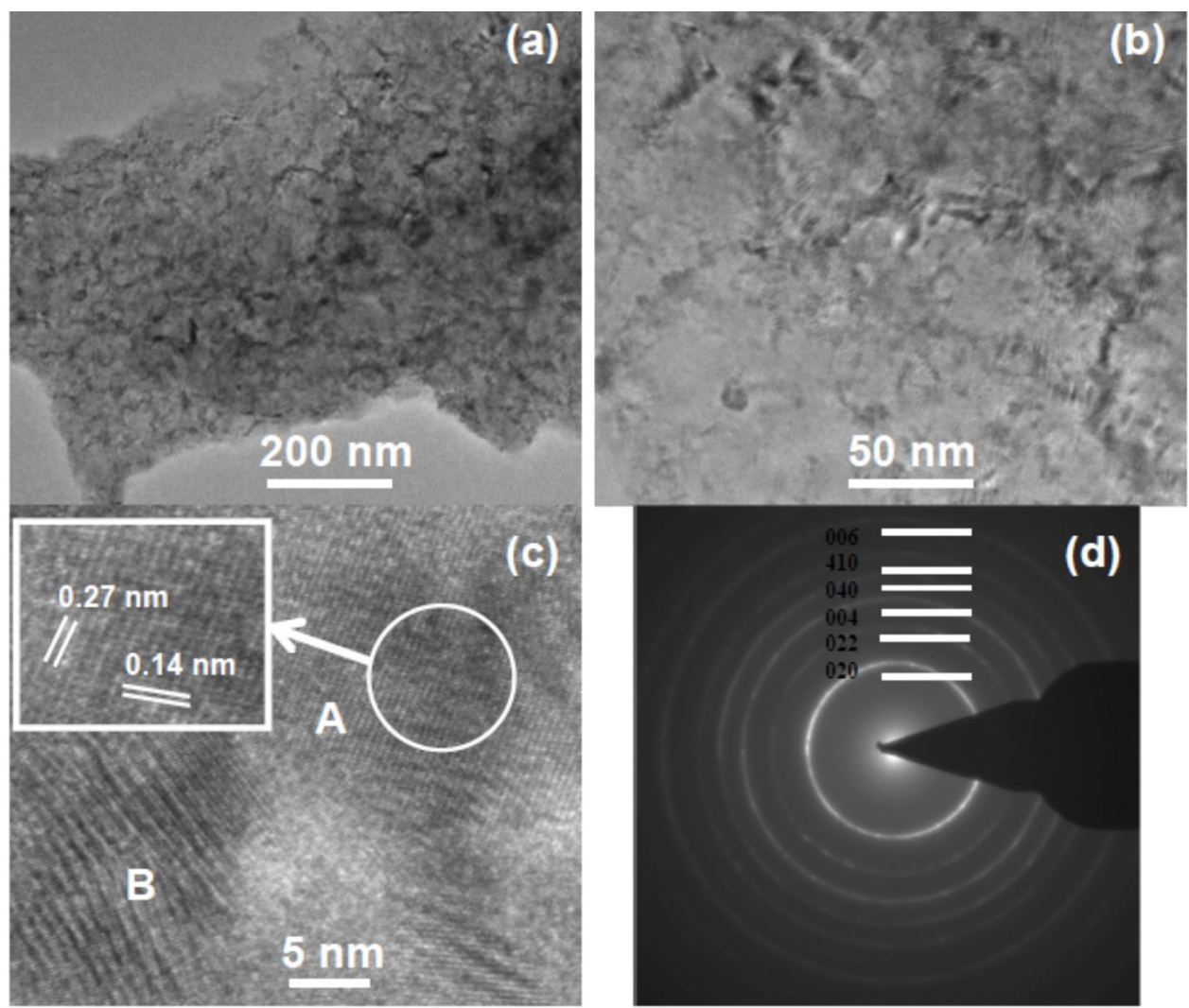

Fig. 13